\documentclass[preprint]{jpsj3}
\usepackage{txfonts}
\usepackage{color}

\begin{document}

\title{Magnetic-Field Dependence of Novel Gap Behavior Related to the Quantum-Size Effect}

\author{Tomonori Okuno$^1$\thanks{okuno.tomonori.77s@st.kyoto-u.ac.jp}, Yuta Kinoshita$^{1}$, Satoshi Matsuzaki$^{1}$, Shunsaku Kitagawa$^1$, Kenji Ishida\thanks{kishida@scphys.kyoto-u.ac.jp}$^1$, Michihiro Hirata$^2$, Takahiko Sasaki$^2$, Kohei Kusada$^3$, Hiroshi Kitagawa$^3$}

\inst{$^1$ Department of Physics, Graduate School of Science, Kyoto University, 606-8502 Kyoto, Japan\\
$^2$ Institute for Materials Research, Tohoku University, 980-8577 Sendai, Japan \\
$^3$ Department of Chemistry, Graduate School of Science, Kyoto University, 606-8502 Kyoto, Japan}

\abst{$^{195}$Pt-NMR measurements of Pt nanoparticles with a mean diameter of 4.0 nm were performed in a high magnetic field of approximately $\mu_0 H = 23.3$ T to investigate the low-temperature electronic state of the nanoparticles.
The characteristic temperature $T^*$, below which the nuclear spin-lattice relaxation rate $1/T_1$ deviates from the relaxation rate of the bulk, shows a magnetic-field dependence.
This dependence supports the theoretical prediction of the appearance of discrete energy levels.}

\maketitle

Since the theoretical prediction of the quantum-size effect (QSE) on metallic nanoparticles by Kubo~\cite{Kubo1962}, intensive theoretical studies on this effect have been conducted~\cite{gorkov1965, Sone1977, Halperin1986}.
Kubo predicted that the electronic properties of metallic nanoparticles will differ from those of the bulk with decreasing particle size and will depend on the parity of the even or odd electron number.
The key point of this theoretical prediction was consideration of the Hamiltonian of the nanoparticles as a random matrix resulting from the inhomogeneous boundary conditions that should be taken into account.
Thus, the energy gap $\Delta$ of each nanoparticle is a probability variable with a certain distribution with a mean value $\delta_\mathrm{Kubo}$.
Later, it was shown that the distribution is not unique but depends on the symmetry of the Hamiltonian~\cite{gorkov1965}.
The distributions generally differ depending on the strength of the magnetic field~\cite{Nomura_Al, Kobayashi_Al}.

We have previously investigated the electronic properties of Pt nanoparticles with nuclear magnetic resonance (NMR) measurements~\cite{Okuno}.
In our measurements, the NMR signals corresponding to the surface could be distinguished from those corresponding to the interior regions of the nanoparticles. This was possible because of the advantage of the large Pt Knight shift owing to the large density of states of $d$ electrons\cite{Clogston1964}.
We found that the nuclear spin-lattice relaxation rate $1/T_1$ of the nanoparticles deviated from the bulk behavior and started increasing below the characteristic temperature $T^*$. The systematic particle size dependence suggests that $k_\mathrm{B} T^*$ would correspond to the ``Kubo'' discrete-energy gap~\cite{Okuno}.
The $H$ dependence of $T^*$ is crucial to support the above hypothesis.
If the above statements are true, $T^*$ will be affected by $H$ via the Zeeman effect.
 In nanoparticles with odd (even) number of electrons the energy gap, or $k_\mathrm{B} T^*$, will increase (decrease) with increasing $H$.
When the Zeeman energy exceeds the gap size, the $H$ dependence will be switched, i.e., in nanoparticles with odd (even) number of electrons, the energy gap, or $k_\mathrm{B} T^*$, will increase (decrease) with increasing $H$.

In this paper, we investigate the $H$ dependence of $T$ and report that $T^*$ shows a non-monotonic dependence on $H$, which supports the realization of discrete energy levels corresponding to the ``Kubo'' gap.

In our previous study~\cite{Okuno}, we focused on the nanoparticles of $d$-electron metals with a large DOS at the Fermi level, which is useful to distinguish the QSE from the surface effect.
In addition, the Pt nucleus is suitable for NMR measurements~\cite{IUPAC}; the gyromagnetic ratio of $^{195}$Pt is large ($\gamma_\mathrm{n}=9.153$ MHz/T) with a nuclear spin of $I=1/2$, and thus, the nuclear quadrupole interaction is absent in the Pt nucleus.
Therefore, the shift and linewidth of the $^{195}$Pt-NMR spectrum are determined based on only magnetic interaction.

We carried out measurements on Pt nanoparticle samples with a mean diameter of 4.0 nm, prepared by the reduction of metal ions.
The sample synthesis conditions are described in the previous paper~\cite{Okuno}.
For the NMR measurements, we used 500-mg-weight samples (Pt nanoparticles coated with Polyvinylpyrrolidone).
The NMR measurements were performed in a wide magnetic field ranging from $\sim$ 2.8 T (25.35 MHz) to $\sim$ 23.3 T (212.03 MHz).
$T_1$ was measured using the saturation-recovery method at each position of the spectrum i.e., at various magnetic fields.
Single-component $T_1$ was evaluated through exponential fitting in the high-temperature range where the Korringa relation holds (metallic range).
At low temperatures, the recovery of the nuclear magnetization shows a multi-exponential behavior; thus, the fitting was performed in two time regions, and only the fastest components are shown in this paper.

\begin{figure}[tb]
\includegraphics[clip, width= 8.6 cm]{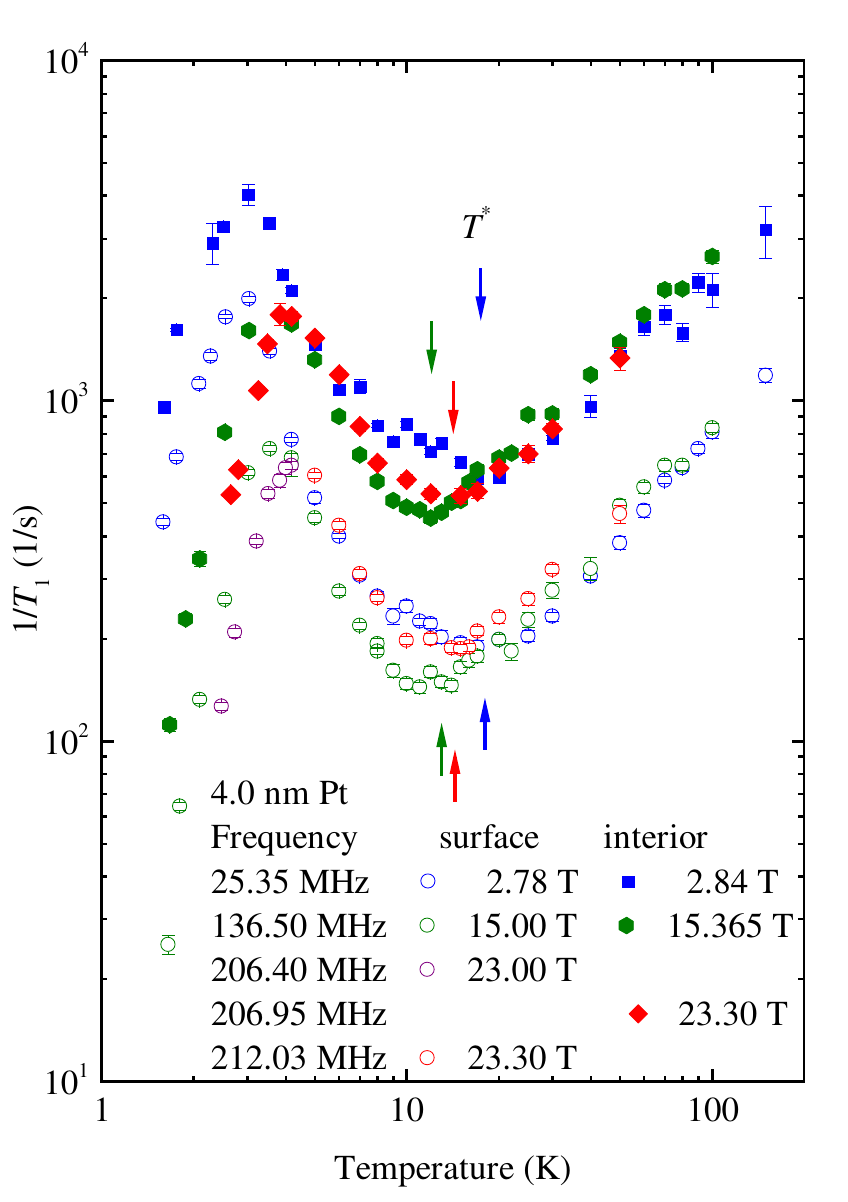}
\caption{\label{fig: T1 field dependence}
(Color online) Temperature dependence of the nuclear spin-lattice relaxation rate $1/T_1$ measured in various magnetic fields at both the surface and the interior positions.
Arrows show the characteristic temperature $T^*$.}
\end{figure}

\begin{figure*}[tb]
\includegraphics[clip, width= 17.2 cm]{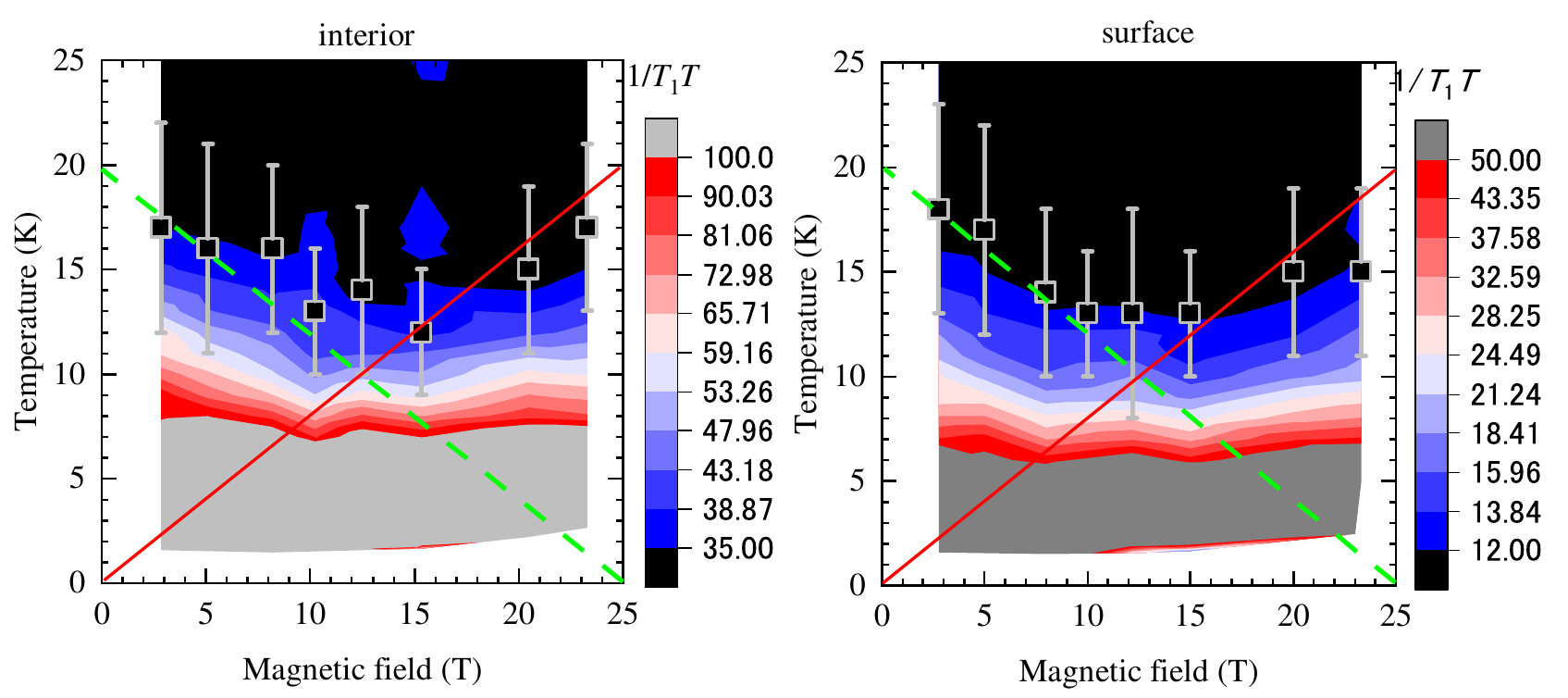}
\caption{\label{fig: gap field dependence}
(Color online) Points: magnetic-field dependence of $T^*$ determined in Fig.~1. Temperature and magnetic-field dependence of $1/T_1T$ are shown by the contour plot.
Dotted line: $T^* = \Delta/k_\mathrm{B} - g\mu_\mathrm{B}\mu_0 H/k_\mathrm{B}.$; Solid line: $T^* = g\mu_\mathrm{B}\mu_0 H/k_\mathrm{B}$, Zeeman splitting model with $\Delta/k_\mathrm{B}=20$ K and $g=1.2$.
}
\end{figure*}
The NMR spectrum did not change in the measured $H$ range.
Therefore, the NMR signals arising from the surface and interior regions were determined based on the Knight-shift values, as done in the previous study\cite{Okuno}.
The temperature dependence of $1/T_1$ on the surface and in the interior regions of the nanoparticles in various magnetic fields is shown in Fig.~\ref{fig: T1 field dependence}.
At high temperatures above 20 K that represented the metallic range in our previous study~\cite{Okuno}, $1/T_1$ was unchanged despite increasing $H$; and the bulk shows a similar behavior. This is consistent with the theoretical prediction that the Kubo effect appears only at low temperatures below the gap size.
On cooling, $1/T_1$ deviated from its behavior in the metallic range and started rapidly increasing below the characteristic temperature $T^*$ and reached a peak, which was followed by a sharp decrease.
With increasing $H$, $T^*$ evidently shifted to a lower temperature and subsequently increased again. 
The peak temperature increased monotonically with the magnetic field, and the peak value of $1/T_1$ continued to decrease.
The magnetic-field dependence of the characteristic temperature $T^*$ is shown with a contour plot of $1/T_1T$ in Fig.~\ref{fig: gap field dependence}.
The $H$ dependence of $T^*$ was observed even for $1/T_1T$.
The experimentally obtained $T^*$ in the interior or the surface regions could be fitted to the Zeeman splitting model of
\begin{align*}
  T^* =
  \begin{cases}
    \Delta/k_\mathrm{B} - g\mu_\mathrm{B}\mu_0 H/k_\mathrm{B}
      & \mu_0 H < \Delta/2g\mu_\mathrm{B} 
      \\
    g\mu_\mathrm{B}\mu_0 H/k_\mathrm{B}
      & \mu_0 H > \Delta/2g\mu_\mathrm{B} 
  \end{cases}
\end{align*}
with $\Delta/k_\mathrm{B} = 20$ K and $g = 1.2$.
Interestingly, the $H$ dependences of $T^*$ and $1/T_1T$ changed at $H$ corresponding to $k_{\rm B} T^*/2$, suggesting that the dominant relaxation mechanism was altered at this field. 
We also note that the value $g=1.2$ estimated from the $H$ dependence is almost the same as the Lande's $g$ factor $g_J=1.2$ of Pt, the electronic configuration of which is [Xe](4$f$)$^{14}$(5$d$)$^9$(6$s$)$^1$.
This indicates the strong spin orbital interaction (SOI) of the Pt nanoparticles.

It was predicted that the Knight shift shows oscillatory dependence against $H$~\cite{gorkov1965}, but such a behavior was not observed in this study.
As described previously\cite{Okuno}, a large SOI can  suppress the QSE in static susceptibility~\cite{Sone1977}. 
Since the Zeeman energy was almost the same as the gap size in Pt nanpparticles, the magnetic-field effects on the Knight shift are considered to be negligibly small.
Other cases can be explained as follows.
In case of Al nanoparticles, for which the gap size exceeds SOI, it was reported that the Knight shift depends on the magnetic field~\cite{Nomura_Al, Kobayashi_Al}.
The Knight shift of the Al nanoparticles decreased compared to that of the bulk at a low temperature and low field, and recovered to the value of the bulk at a high field, nearly corresponding to the gap size.
The gap size of Cu nanoparticles~\cite{Tunstall1991, Tunstall1994} is extremely large and is not affected by magnetic field.

In summary, we performed high-field NMR measurements of Pt nanoparticles with a mean diameter of 4.0 nm to investigate the electronic states of the $d$-electron system nanoparticles.
We found that the behavior of $T^*$ related to the QSE shows $H$ dependence for both the surface and interior regions, which can be reasonably explained in terms of the appearance of discrete energy levels, as predicted by Kubo.

\subsection*{Acknowledgments}
The authors would like to thank K. Kinjo, G. Nakamine, A. Ikeda, T. Taniguchi, S. Yonezawa, Y. Maeno, and M. Koyama for valuable discussions and comments.
This work was partially supported by the Kyoto University LTM Center and
Grant-in-Aids for Scientific Research (KAKENHI) (Grant Nos. JP15H05882, JP15H05884, JP15K21732, JP15H05745, JP19K14657, and JP19H04696).

\end{document}